 \documentclass[12pt]{article}

\def\fun#1#2{\lower3.6pt\vbox{\baselineskip0pt\lineskip.9pt
        \ialign{$\mathsurround=0pt#1\hfill##\hfil$\crcr#2\crcr\sim\crcr}}}

\renewcommand\({\left(}
\renewcommand\){\right)}

\newcommand\eq[1]{Eq.~(\ref{#1})}
\newcommand\eqs[2]{Eqs.~(\ref{#1}) and (\ref{#2})}
\newcommand\eqss[3]{Eqs.~(\ref{#1}), (\ref{#2}) and (\ref{#3})}

\newcommand\ee{\end{equation}}
\newcommand\be{\begin{equation}}
\newcommand\eea{\end{eqnarray}}
\newcommand\bea{\begin{eqnarray}}

\newcommand\GeV{\,\mbox{GeV}}

\newcommand\mpl{M_{\rm P}}

\newcommand{\lsim}{\mbox{\raisebox{-.9ex}{~$\stackrel{\mbox{$<$}}{\sim}$~}}}

\newcommand{\gsim}{\mbox{\raisebox{-.9ex}{~$\stackrel{\mbox{$>$}}{\sim}$~}}}

\newcommand{\lgeq}{\mbox{\raisebox{-.1ex}{~$\stackrel{>}{_<}$~}}}

\def\dslash{\not{\hbox{\kern-2pt $\partial$}}}
\def\Dslash{\not{\hbox{\kern-4pt $D$}}}
\def\Oslash{\not{\hbox{\kern-4pt $O$}}}
\def\Qslash{\not{\hbox{\kern-4pt $Q$}}}
\def\pslash{\not{\hbox{\kern-2.3pt $p$}}}
\def\kslash{\not{\hbox{\kern-2.3pt $k$}}}
\def\qslash{\not{\hbox{\kern-2.3pt $q$}}}

 \newtoks\slashfraction
 \slashfraction={.13}
 \def\slash#1{\setbox0\hbox{$ #1 $}
 \setbox0\hbox to \the\slashfraction\wd0{\hss \box0}/\box0 }

\def\ee{\end{equation}}
\def\be{\begin{equation}}

\def\calp{{\cal P}}

\newcommand\sub[1]{_{\rm #1}}

\begin{document}

\newcommand\sigmaosc{_{\sigma\,\rm osc}}


\begin{center}
{\Large\bf The Curvaton as a Pseudo-Nambu-Goldstone Boson}

\bigskip
\bigskip

{\Large   K. Dimopoulos$^{*,\dag}$, 
D.H.~Lyth$^*$, A. Notari$^{**}$  and A. Riotto$^{***}$}

\bigskip

$^*${\it Physics Department, Lancaster University, Lancaster LA1 4YB,
UK}

$^\dag${\it Department of Physics, University of Oxford, 
Keble Road, Oxford OX1 3RH, UK}

$^{**}${\it Scuola Normale Superiore, Piazza dei Cavalieri 7, Pisa I-56125, 
Italy}

$^{***}${\it INFN, Sezione di Padova, via Marzolo 8, I-35131,  
Italy}


\vskip 2cm
\begin{abstract}
\noindent
The field responsible for the cosmological curvature perturbations
generated during a stage of primordial inflation might be the 
 ``curvaton'', a field different from the inflaton field.
To keep  the effective mass of the curvaton small enough compared to
the Hubble rate during inflation one may not invoke 
supersymmetry since the latter  is broken 
by the vacuum energy density. In this paper we 
propose the idea that the curvaton 
is a pseudo Nambu-Goldstone
boson (PNGB) so that its potential and  mass   vanish in the
limit of unbroken symmetry.  We give a general framework within
which PNGB curvaton candidates should be explored. Then we explore
various possibilities, including the case where
the curvaton can be
identified with the extra-component of a gauge field
in a compactified five-dimensional theory (a Wilson line), where it 
comes  from a Little-Higgs
mechanism, and where it
is a string axion so that supersymmetry is essential.

\end{abstract}

\vskip 1cm 

\begin{flushleft}
DFPD-TH/03/12
\end{flushleft}

\end{center}
\newpage

\section{Introduction}

\noindent
It is now clear that structure in the Universe 
comes primarily from an almost scale-invariant superhorizon curvature 
perturbation \cite{treview}.
This perturbation originates presumably from the vacuum fluctuation,
 during almost-exponential inflation,  of some field with mass
much less than the Hubble parameter $H$. Indeed, every
such field  acquires a nearly scale-invariant classical perturbation.
The question is, which field is responsible for the curvature perturbation?

With two exceptions \cite{sylvia,lm}, the universal assumption until
2001 was that the responsible field was the inflaton, defined in this
context as the field whose value determines the end of inflation.
Then it was proposed instead \cite{lw}
that the responsible field is  some
 `curvaton' 'field different from the inflaton.
The curvaton field oscillates during some radiation-dominated era,
causing its energy density to grow to at least 1\% of the total
and thereby generating the curvature perturbation.
The   curvaton hypothesis has since
been the subject of a lot
of attention
 \cite{mt1,andrew,mt2,fy,luw,hmy,hofmann,dl,bck1,bck2,%
mormur,ekm,mwu,postma,fl,gl,kostasquint,giov,lu,omit,dllr1,ejkm,%
03dl,dllr2} because it opens up new possibilities both for
theory and observation.\footnote
{In a variant of the curvaton scenario \cite{bck1,bck2}
the  curvaton field only indirectly causes the growth of the 
curvature perturbation. Recently, a completely different
scheme has
been proposed \cite{dgz,kofman} in which the responsible field
acts by perturbing the inflaton decay rate.}

The requirement that the effective curvaton  mass be much less than the Hubble
parameter during inflation is a severe constraint. In this respect
the situation for the curvaton is the same as that for the inflaton
in the inflaton scenario\footnote
{In the curvaton scenario, the effective inflaton mass during inflation
might be of order the Hubble parameter
\cite{dl}.}.
In contrast with the inflaton case though, the effective
curvaton  mass should
preferably also be much less than $H$ even after inflation, since otherwise
the curvaton perturbation has non-trivial evolution and one has to be careful
that its perturbation is not  driven to zero \cite{dllr1,dllr2}.

To keep  the effective mass of the  inflaton or curvaton small enough,
it seems natural to invoke supersymmetry (SUSY). 
Flat directions are frequent in supersymmetric
theories, and they are stable under radiative corrections, as long as
SUSY is not broken.  However, in the early Universe SUSY is broken
by the energy density, and as a result the effective mass-squared
of each scalar field generically receives a contribution of order
$\pm H^2$,  both during \cite{dfn,chrr,cllsw} and after \cite{drt,pre} 
inflation.

An alternative possibility for keeping the effective mass sufficiently small
is to make the inflaton or the curvaton a pseudo Nambu-Goldstone
boson (PNGB), so that its potential vanishes  in the limit
where  the  corresponding global  symmetry is unbroken.
Then the effective mass-squared of the curvaton or the inflaton
(defined as the second derivative
of its  potential) vanishes in the
limit of unbroken symmetry and can indeed be kept small by keeping the breaking
sufficiently small.

However, in
 the case of the inflaton the vanishing of its  potential
in the limit of unbroken symmetry 
raises two complications.
First, one has to ensure that the entire scalar field potential does not
vanish in the limit of unbroken symmetry. This means that one has to work
with a hybrid inflation model, where 
 the potential of some `waterfall' field different
from the inflaton is nonzero during inflation.
Second, one has to ensure  that the slope  $V'$  of the inflaton
potential is not too small during inflation, 
since otherwise the curvature perturbation is too big.
In particular, in the regime $V'\lsim H^3$ the quantum fluctuation
of the inflaton field would dominate its classical motion, leading to
a curvature perturbation
 of order one (`eternal' inflation).
For these and other reasons, it turns out that
 models making the inflaton a PNGB are 
fairly complicated,  whether in the original context
of the  symmetry associated with a supergravity potential of the 
no-scale type \cite{cllsw,ewan,lmg,cr} or in the more recent context 
\cite{subir,cs,newwilson1,newwilson2,newwilson3} of 
using a   symmetry such as $U(1)$, $SU(N)$ or $SO(N)$.

In contrast, the vanishing of the curvaton potential in the limit
of exact symmetry raises no problems at all.
In particular, it is quite all right for the slope of the curvaton
potential to be
so small that the entire range of the curvaton field lies in the
quantum regime \cite{lw}. In this paper we explore in detail the
idea that  the curvaton is a PNGB. We consider the simplest case
that the symmetry is a $U(1)$, and look at   schemes 
where SUSY is optional as well as
 the string axion case where it is essential.
 The layout of the paper is as follows. In Section \ref{general} we establish
some general formulas. In Section \ref{nonrenorm} we recall the case
\cite{lw} that the symmetry is broken by a non-renormalizable term.
In Section \ref{wilson} we consider the case that the symmetry
is broken by a Wilson line in an extra dimension, and in Section
\ref{lh} we consider the 
little-Higgs
mechanism. (These cases for the inflaton have recently been considered
in references  \cite{newwilson1,newwilson2,newwilson3}.) 
In Section \ref{axion} we consider
the case that the curvaton is a string axion, and we 
conclude in Section \ref{conclusion}.

\section{General estimates}

\label{general} 

\subsection{The curvaton potential}

\noindent
We assume that the curvaton field $\sigma$ is the PNGB
of a broken $U(1)$ symmetry, with some finite range
$-\pi v<\sigma < \pi v$. Without loss of generality we can take
$\sigma=0$ to be the minimum of the potential. Keeping only 
one term in the power series the potential is then
\bea
\label{form}
V(\sigma) &=& (vm)^2 \left(1-\cos \(\frac{\sigma}{v} \)\right) \label{vpot}\\
& \simeq  & \frac12 m^2 \sigma^2
\label{vpot2}
\,,
\eea
where the second line is valid
for small $\sigma\ll v$ so that $m$ is the curvaton mass.
Additional terms in the power series will modify the potential
in some region around the maximum, but we assume that this region is
small enough that it can be ignored.  Because we are dealing with a PNGB,
we  shall in general assume that this potential is not significantly
modified in the early Universe.


The curvaton scenario assumes almost-exponential inflation, with some
Hubble parameter $H_*$. To avoid excessive gravitational wave production
one needs $H_*\lsim 10^{-5}\mpl$.
 We shall go further, and assume
that inflation involves  a slowly-rolling inflaton field. Then the requirement
that the inflaton contribution to the curvature perturbation is some fraction
$x$ of the total 
gives \cite{dl}
\be
H_*/\mpl \lsim  10^{-5}x
\label{hbound1}
\,.
\ee
The  curvaton scenario requires $x\ll 1$.
The requirement that the 
curvaton potential be negligible during inflation corresponds to
\be
v m \ll \mpl H_*
\,.  \label{potential}
\ee
It is assumed that the curvaton is light during inflation,
\be
m\ll H_*
\label{hbound2}
\,,
\ee
so that  on super-horizon scales it has a classical
perturbation with an almost  flat spectrum given by
\be
\calp_{\delta\sigma}^\frac12 = H_*/2\pi
\,.
\ee


\subsection{The curvature perturbation}


After horizon exit, the unperturbed curvaton field $\sigma$ and its
perturbation $\delta\sigma$ evolve according to the classical equations
\bea
\ddot \sigma + 3H\dot\sigma + V'(\sigma) &=& 0 \\
\ddot {\delta\sigma} + 3H\dot{\delta \sigma} + V''(\sigma)\delta\sigma 
&=& 0 
\,,
\eea
where the primes denote derivatives with respect to $\sigma$.
Because of the flatness condition \eq{hbound2} on the potential,
the   changes in $\sigma$ and $\delta \sigma$ are  negligible until 
$H\sim m$. 

When $H\sim m$, 
 the field 
starts to  oscillate around zero. At this stage the 
 curvaton energy density is
$\rho_\sigma = \frac12m^2 \sigma_*^2$ while the total is
$\rho\sim H^2\mpl^2$. The fraction is therefore
\be
\left. \frac{\rho_\sigma}{\rho} \right|_{H=m}
 \sim \( \frac{\sigma_*}{\mpl} \)^2
\,
\label{rsigfrac}
\ee
The fraction is small provided that $\sigma_*\ll \mpl$, and we shall
see in a moment that this condition is equivalent to the requirement
that the inflaton gives only a small contribution to the 
curvature perturbation.

After a few Hubble times the oscillation will be sinusoidal
except for the Hubble damping. The  energy density $\rho_\sigma$
will then be proportional to the square of the oscillation amplitude,
and will scale like the inverse of the locally-defined comoving volume
corresponding to matter domination. On the spatially flat slicing, 
corresponding to uniform local expansion, its perturbation has a constant
value
\be
\frac{\delta\rho_\sigma}{\rho_\sigma} = 2q \(\frac{\delta\sigma}{\sigma} \)_*
\,.
\ee
The  factor $q$  accounts for the evolution of the field
from the time that $m/H$ becomes significant, and will be
  close to 1 provided that  $\sigma_*$ is not too close
to  the maximum value $\pi v$.\footnote
{The correction factor is estimated in \cite{myaxion}. 
The factor $q$ in principle also includes evolution of the field before
the oscillation starts, but as already mentioned this evolution will
be negligible when the curvaton is a PNGB. If instead the curvaton
is not a PNGB, $q$  may  be  very different from 1 \cite{dllr1,dllr2}}

The curvature perturbation $\zeta$ is supposed to be negligible when the
curvaton starts to oscillate, growing during some radiation-dominated
era when $\rho_\sigma/\rho\propto a$.
After the curvaton decays $\zeta$  becomes constant. In 
the approximation that the curvaton decays instantly
(and setting $q=1$) it is then given by
\cite{luw}
\be
\zeta \simeq \frac{2r}3 \( \frac{\delta \sigma}{\sigma} \)_*
\label{zetapred}
\,,
\ee
where 
\be
r\equiv \left. \frac{\rho_\sigma}{\rho} \right|\sub{DEC}
\label{rdef}
\,,
\ee
and the subscript DEC denotes the epoch of decay. The corresponding spectrum
is 
\be
\calp_\zeta^\frac12 \simeq  \frac{2r}3  \frac{H_*}{2\pi \sigma_*}
\label{specpred}
\,,
\ee
It  must
match the observed  value \cite{LLbook,wmapparam}
$\calp_\zeta^\frac12 = 5\times
10^{-5}$, which means that
\be
\frac{H_*}{\sigma_*} \simeq 5 \times 10^{-4} / r
\label{hoversigma}
\,.
\ee

The current
WMAP bound on non-gaussianity \cite{wmapnong} requires $r>9\times 10^{-3}$.
Using \eqs{hbound1}{hoversigma} we find 
\be
\frac{\sigma_*}{\mpl} \simeq  10^{-2} r^{-1} x \lsim x
\,.
\ee
As advertised, this is small because the inflaton is supposed to 
contribute  a negligible fraction of the observed curvature perturbation.

\subsection{The epoch of curvaton decay}

The curvaton scenario can lead to non-gaussianity,  and also to 
isocurvature perturbations which are generated by the curvaton field
and hence are fully correlated  with the curvature perturbation
(residual isocurvature perturbations). Whether these are possible
depends on the fraction $r$, that the curvaton 
contributes to the energy density by the time that it decays.
The regime  $r<10^{-2}$ is forbidden, because it gives
non-gaussianity above the level permitted by
present WMAP data. In the regime $10^{-2} < r \ll 1$ there is significant
 non-gaussianity, which  can be detected by
by the PLANCK satellite or even by future WMAP data. 
On the other hand, isocurvature
matter  perturbations in this regime are either zero (if the matter is created
after curvaton decay) or else bigger than is allowed by observation.

In the regime $r\simeq 1$ the situation is reversed.
Non-gaussianity will be only at the second-order level, as in the inflaton 
scenario \cite{nongauss}, and difficult to observe in the foreseeable
future. On the other hand, isocurvature matter density perturbations
can be generated at a level that will be  detectable in the foreseeable
future.

In order to evaluate  $r$, one needs the decay rate $\Gamma$ of the 
curvaton which determines the epoch
$H\sim \Gamma$ of  curvaton
decay. 
One also needs to know 
 the behaviour of the scale
factor  while  the curvaton is oscillating.
Here, we shall consider the two simplest cases. 
First, that the non-curvaton energy density corresponds to radiation
throughout the oscillation. Second, that
 the non-curvaton energy density corresponds to matter
until some
`reheating' epoch, after which it again corresponds to radiation.
(In the simplest version of this second case,
 the matter domination era prior to the `reheating' epoch
will begin right after inflation.)
In both cases, we use  \eq{rsigfrac},
along with
the fact that matter energy density (including the energy density of the 
oscillating curvaton) scales like $a^{-3}$ while radiation energy density 
scales like $a^{-4}$, with the scale factor being
$a\propto t^\frac23$ during matter domination and
$a\propto t^\frac12$ during radiation domination. 
We find that 
at the decay epoch, the fraction $r$ of the energy density
due to the curvaton is given by
\be
\frac{m\sigma_*^4}{\Gamma \mpl^4}\min\left\{1,\frac{H\sub{REH}}{m}\right\}
\sim \left\{
\begin{array}{ll}
r^2 &   \ \ (r\ll 1) \\
 (1-r)^{-\frac32} & \ \ (1-r\ll 1) 
\end{array}
\right.
\,,
\label{r1}
\label{curvfrac}
\ee
where $H\sub{REH}$ is the Hubble parameter at reheating.
Using \eq{hoversigma} and $r<1$ this leads to the bound

\be
\frac{H_*}{\mpl} \gsim  5\times 10^{-4} \( \frac \Gamma m \)^\frac14 
\label{hstarbound}
\,.
\ee



Since  the curvaton should not disturb big bang nucleosynthesis (BBN)
we require that its decay occurs earlier, which introduces the bound
\begin{equation}
\Gamma>H_{\rm BBN}=\frac{\pi}{3}\left(\frac{g_{\rm BBN}}{10}\right)^{1/2}
\frac{T_{\rm BBN}^2}{M_{\rm P}}\simeq 4.5\times 10^{-25}{\rm GeV},
\label{bbn1}
\end{equation}
where \mbox{$T_{\rm BBN}\simeq 1$ MeV} and \mbox{$g_{\rm BBN}=10.75$} is
the relativistic degrees of freedom at the time. If the decay rate
is of gravitational strength, corresponding to $\Gamma \sim m^3/\mpl^2$, 
 the above bound demands \mbox{$m>10$ TeV}.

\subsection{The assumption of  randomization}

\label{random}

In the   curvaton scenario,  the curvature
perturbation $\zeta$ is given up to possible post-inflationary evolution
by \eq{zetapred}, involving
the fractional perturbation $(\delta \sigma/\sigma)_*$
 of the curvaton field a few Hubble times after horizon exit during inflation.
The perturbation $\delta\sigma$  is completely under control, being a 
scale-invariant gaussian quantity with the famous spectrum $(H_*/2\pi)^2$.
To calculate the {\em fractional} perturbation though, one needs the
unperturbed value $\sigma_*$. It represents  the average value
  of $\sigma$ within the comoving region that will become our
 observable Universe, evaluated  a few Hubble times after
that region leaves  the horizon.
The average of $\sigma$,  within 
a region of fixed size $H_*^{-1}$,
evolves 
under the combined effects of the
classical slow roll  and the quantum fluctuation.
In one Hubble time,
the classical slow roll  gives a change $\Delta\phi = -V'/H_*^2$, while
the quantum fluctuation gives a random contribution $\Delta \phi =
\pm H_*/2\pi$.
The classical motion dominates if $|V'|\gg H_*^3$, but  even if this 
condition holds early on it cannot be satisfied indefinitely because
the classical motion of $\sigma$ will always be towards a minimum where
$V'$ vanishes. 

We shall therefore assume\footnote
{In  \cite{dllr2} it was necessary to make the  opposite assumption, in order
to obtain a sufficiently big curvature perturbation.}
 that before 
our Universe leaves the horizon the quantum fluctuation has come to 
dominate,  placing  $\sigma_*$ within a region around the minimum given by
$|V'| \lsim H_*^3$. 
Remembering that the maximum value of $\sigma$ is $\pi v$, and 
using  the approximation $V\simeq \frac12 m^2 \sigma^2$, one can
see that this region corresponds to
\be
\sigma_* \lsim \min \left\{ H_*^3/m^2, v \right\}
\label{sigmaexp}
\,.
\ee

{}From these considerations, we see that 
 there is no precise prediction for the spectrum of the
observed curvature perturbation
in the curvaton scenario.  Rather,  the spectrum 
in  the curvaton scenario  is itself a stochastic quantity, depending on 
the location of our Universe within the much larger perturbed region
that presumably surrounds it. This state of affairs arises because
the spectrum, as far as observation is concerned, should not be defined
within an arbitrarily large region around us. Instead, it should
be  defined in a region around us 
 only a few orders of magnitude bigger than the one we observe.
Up to a numerical factor of order unity, the spectrum is simply
the mean-square value of the curvature perturbation is such a
region.\footnote
{An analogous  situation is encountered for the axion isocurvature
perturbation irrespective of the scenario for the generation of the
curvature perturbation \cite{myaxion}.} 

If it is assumed that we live in a typical region, then the
order of magnitude of $\sigma_*$ is predicted to be the
right hand side of \eq{sigmaexp}.
 In that case,
the  order of magnitude of the value  of the spectrum (given by
\eq{specpred}) is also predicted. The spectrum in other locations
is then typically about the same as in the our region, and is otherwise
much {\em bigger}.

If instead it is assumed that we live in a  region 
where $\sigma_*$ is many orders of magnitude smaller than the typical
value, then there are regions where the spectrum of the curvature perturbation
is both much bigger and much smaller. In that case, the fact that
the spectrum has roughly the observed value would presumably  be the  result
of anthropic selection \cite{martin}. Which of these two assumptions is 
observationally viable depends on the model.


\subsection{Inequalities following from the randomization assumption}

Using \eq{hoversigma}, one can write \eq{sigmaexp} as
\be
(r/A)^\frac12 m \lsim H_* \lsim (A/r) v
\,,
\label{hbound3}
\ee
where $A=5\times 10^{-4}$. The consistency of the two inequalities
 requires 
\be
m/v  \lsim 10^{-2} \( \frac{10^{-2}}{r} \)^\frac32 
< 10^{-2}
\label{moverv1}
\,.
\ee

If we live in a typical location, $H_*$ will approximately 
saturate  either the 
upper or the lower limit of \eq{hbound3}. Allowing $r$ to vary over its
range $10^{-2} \lsim r< 1$ merges these possibilities if 
\be
m/v \gsim  A^\frac32 \sim 10^{-5}
\,.
\ee
Otherwise they represent two distinct ranges of $H_*$, with a gap between 
them. 

{}The first inequality of \eq{hbound3} 
leads to 
\be
vm \lsim 
0.2 v H_*
\,.
\ee
Provided that $v\lsim \mpl$, this
 guarantees 
that the curvaton potential is negligible during
inflation (\eq{potential}). 

Another way of writing the first inequality is
\be
\frac{m^2}{H_*^2} \lsim    0.05 \( \frac{10^{-2}}{r} \)
< 0.05
\label{mbound}
\,.
\ee
This is particularly interesting, because the spectral index in the curvaton
scenario is given by \cite{lw}
\be
n=1 + \frac23\frac{V''(\sigma_*)}{H_*^2} - 2\epsilon_H
\,,
\label{npred}
\ee
where $\epsilon_H\equiv |\dot H/H^2|_*$. It is expected that PLANCK
 will measure $n$ to an accuracy $\pm 0.01$, which is
therefore the  target accuracy for theoretical predictions. 
Almost all inflation models give $\epsilon_H\ll 0.01$ so that this term
can be ignored, and \eq{mbound} shows that the other term is also 
negligible  unless $r$ is close to saturating its present bound.
We conclude that 
{\em the spectral index is almost certainly indistinguishable from $1$
if the curvaton is a PNGB}. This is line with  the  general expectation
for the curvaton scenario \cite{dl}, coming from the fact that
in contrast with the inflaton, 
the curvaton does not know about the end of inflation.

Finally, we note that
the second inequality of \eq{hbound3} with \eq{hstarbound}
leads to
\be
\frac v \mpl \gsim 10^{-1} \( \frac \Gamma m \)^\frac14
\label{voverplanck1}
\,.
\ee
In a particular model, one should look out for a possible conflict between
this bound and \eq{moverv}.


\subsection{Implementing the estimates}

For a  PNGB to be  a viable curvaton candidate, its parameters must
satisfy the inequalities \eqss{bbn1}{moverv1}{voverplanck1}, 
which we repeat here for convenience:
\bea
\Gamma &\gsim& 4.5\times 10^{-25}{\rm GeV} \label{bbn} \\
m/v  &\lsim&  10^{-2} \( \frac{10^{-2}}{r} \)^\frac32 
< 10^{-2} \label{moverv} \\
\frac v \mpl & \gsim&  10^{-1} \( \frac \Gamma m \)^\frac14
\label{voverplanck}
\,.
\eea
In each case, we will check that these requirements can be satisfied.

In addition, the scale of inflation $H_*$ must lie within the range
\eq{hbound3} and must satisfy \eq{hstarbound}. We consider these bounds
though only if they are of particular interest for the curvaton candidate
itself. The same goes for \eq{curvfrac}, which in principle
determines the parameter $r$ if we know enough about the cosmology.



\section{The symmetry broken by non-renormalizable terms}

\label{nonrenorm}
\noindent
A simple possibility, mentioned already in \cite{lw},
 is that the symmetry is broken mainly  by non-renormalizable
terms. Such a situation has been widely discussed in the case of the 
Peccei-Quinn symmetry, 
where it would be a disaster since the breaking in that case 
is suppose to be through QCD instantons.
Following \cite{lw}, suppose that only one field $\Psi$ 
spontaneously breaks the symmetry, and only a single non-renormalizable
term explicitly breaks it. Keeping only that term,
 the potential of $\Psi$ is
\be
V(\Sigma) = V_0 - m_\Sigma^2 |\Sigma|^2 + \lambda
|\Sigma|^4 + \lambda_d \(  \frac{\Sigma^d}{\mpl^{d-4}} + {\rm c.c.} \)
\,,
\ee 
where 
$\lambda_d\sim 1$ on the assumption that the  ultra-violet cutoff
is at the Planck scale. 
This generates a VEV $\langle |\Psi| \rangle\equiv \sqrt2 v$, in which 
the canonically-normalized PNGB $\sigma$  defined by
$\Sigma =  \sqrt 2 v e^{i\sigma/v}$ has the potential \eq{vpot}
with
\be
m^2 \sim \mpl^2 \( \frac v \mpl  \)^{d-2}
\,.
\ee
This estimate of $m$ will remain valid if other fields spontaneously break the 
symmetry, provided that their VEV's are of similar magnitude.
Using it, the bound \eq{moverv} becomes
\be
 \frac v\mpl \lsim 10^{-\frac{4}{d-4}}
\,.
\label{vbound2}
\ee


To evaluate the other bound  \eq{voverplanck} we need an estimate
of the curvaton decay rate.
If the decay proceeds through dimensionless couplings 
of $\Sigma$ that are of order 1, the  rate will be
$\Gamma \sim m^3/v^2$. Then  the bound \eq{voverplanck}   becomes
\be
(v/\mpl)^{8-d} > 10^{-4}
\,.
\ee
It contradicts \eq{vbound2}  for $d=5$ and $6$, but is compatible for 
$d=7$ and trivial for higher values.
If instead the curvaton decay is of gravitational
strength, $\Gamma \sim m^3/\mpl^2$ and then \eq{voverplanck} becomes
\be
(v/\mpl)^{6-d} > 10^{-4}
\,.
\ee 
This  contradicts \eq{vbound2}  for $d=5$ and is trivial for higher values.
Subject to these restrictions, this PNGB model seems to satisfy all of the 
requirements.

\section{The curvaton as a Wilson Line}

\label{wilson}
\noindent
In this section we discuss the possibility that  the curvaton is
 identified with the extra-component of a gauge field
in a compactified five-dimensional theory. 
Spurred by similar recent considerations applied to  models
of 
inflation 
\cite{newwilson1,newwilson2,newwilson3} and quintessence \cite{prr},
we 
consider a five-dimensional model with the extra fifth dimension 
compactified on a circle of radius $R$  and identify the
curvaton with the fifth component $A_5$ of an abelian gauge field
$A_M$ ($M=0,1,2,3,5)$
propagating in the bulk (the generalization to the  non-abelian 
case is straightforward). As such, the curvaton field cannot have
a local potential because of the higher-dimensional gauge invariance. 
However, a non-local potential as a function of the 
gauge-invariant Wilson line

\begin{equation}
e^{i \theta}=e^{i\,\oint\,g_5 A_5\,dy}\, ,
\end{equation}
where $y$ is the coordinate along the fifth dimension, $0\leq y< 2\pi R$, 
will be generated in the presence of fields charged under the
abelian symmetry \cite{hosotani}

Writing the field $A_5$ as 
\begin{equation}
A_5=\frac{\theta}{2\pi g_5 R}\, ,
\end{equation}
where $g_5$ is the five-dimensional gauge coupling constant, at energies below
the scale $1/R$, and $\theta$ looks like a four-dimensional field
with Lagrangian

\begin{equation}
{\cal L}=\frac{1}{2 g_4^2(2\pi R)^2}\left(\partial_\mu\theta\right)^2-
V(\theta)\, ,
\label{e2}
\end{equation}
where $g_4=g_5/(2\pi R)^{1/2}$ is the four-dimensional gauge coupling 
constant. The canonically normalized field  is $\sigma=v\theta$, with
\be
v=\frac{1}{2\pi g_4 R}
\label{wlinev}
\,.
\ee

The higher-dimensional nature
of the theory preserves the curvaton potential from acquiring 
dangerous corrections, and non-local effects must be
necessarily exponentially suppressed because the typical length
of five-dimensional quantum gravity effects $\sim M_5^{-1}$, where
$M_5$ is the five-dimensional Planck scale, is much smaller than the
size of the extra-dimensions. The global nature of the 
Wilson line preserves its potential from acquiring large 
ultraviolet (local) corrections of the form $\Lambda_{\rm UV}^2\sigma^2$ -- 
where $\Lambda_{\rm UV}$ is the ultraviolet cut-off of the theory --
and therefore the flatness of the potential is preserved. This makes
the Wilson line a perfect candidate for a curvaton 
field.

Let us now turn to the form of the curvaton 
potential.
We assume that the 
potential for the curvaton  field is generated radiatively by a set
of bulk fields which are charged under the $U(1)$ symmetry with charges
$q_a$. 
>From the 
four-dimensional point of view, this is equivalent to having  a 
tower of Kaluza-Klein states with squared masses 
\begin{equation}
m_a^2=\left(\frac{n}{R}+g_4\, q_a\, \sigma\right)^2\, ,\,\,\,
\,\, (n=0,\pm1,\pm2,\dots)\, .
\end{equation}
Borrowing from finite temperature
field theory calculations, the potential can be
written as \cite{pq}
\begin{equation}
V(\sigma)=\frac{1}{128\pi^6 R^4}\,{\rm Tr}\left[V\left(r^F_a,\sigma\right)
-V\left(r^B_a,\sigma\right)\right]\, ,
\end{equation}
where the trace is over the number of degrees of freedom, 
the superscripts $F$ and $B$ stand for fermions and bosons, 
respectively and
\begin{equation}
\label{pot}
V\left(r_a,\sigma\right)=3\,{\rm Li}_5\left(r_a\right)+{\rm h.c.}\, .
\end{equation}
We have defined 
\begin{equation}
r_a= e^{i q_a \sigma/v}\, ,
\end{equation}
and in Eq. (\ref{pot}) the functions ${\rm Li}_n(z)$ stand for   
the polylogarithm
functions 
\begin{equation}
{\rm Li}_n(z)=\sum_{k=1}^{\infty}\frac{z^k}{k^n}\, .
\end{equation}
The potential (\ref{pot})
is well approximated by the form (\ref{form}) with
\begin{equation}
\label{crucial}
V_0\simeq \frac{3c}{16\pi^6}\frac{1}{R^4}
\,,
\end{equation}
{\sl COMMENT: next 4 paras new}
where $c\sim 1$ is a numerical coefficient depending upon the charges
of the bulk fields. With \eqs{vpot}{wlinev} this corresponds to mass
\be
m\sim 10^{-1} g_4/R
\label{wlinem}
\,.
\ee
The curvaton decays  into the zero
modes of the bulk fields, with rate
\begin{equation}
\Gamma\sim 10^{-1} \,g_4^2\, m\sim 10^{-2}\,\frac{g_4^3}{R}
\label{wlinegamma}
\,.
\end{equation}

The conditions \eqss{bbn}{moverv}{voverplanck} give respectively
\bea
g_4 &\gsim& 10^{-13} (\mpl R)^\frac13 \label{g41}\\
g_4 &\lsim&  10^{-1} \label{g42} \\
g_4 &\lsim& (\mpl R) ^{-\frac23} \label{g43}
\,.
\eea
The first and third of these together lead to the weak constraint
$R^{-1} \gsim 10^5\GeV$. 

Since the  effective four-dimensional
field theory under consideration is supposed to be valid during inflation,
we require $H_* R \ll 1$, and this leads to two more constraints on the
curvaton parameters. One, corresponding to the first bound in \eq{hbound3}
is automatic by virtue of \eq{g42}. The other, corresponding to 
\eq{hstarbound}, is 
\be
g_4 \lsim 10^7 (\mpl R)^{-2}
\label{g44}
\,,
\ee
which with \eq{g41} gives
\be
1/R \gsim 10^{10}\GeV
\label{Rbound}
\,.
\ee

The  constraints follow just from the requirement that
inflation lasts long enough for the curvaton to be somewhere in
 the quantum regime. The most important results are the lower bound \eq{Rbound}
on the size of the extra dimension, and  \eq{g44}
which requires a very  small coupling $g_4$ unless the extra dimension
is quite small.

\section{The curvaton as a Little Higgs}

\label{lh}
\noindent
Little Higgs (LH) theories \cite{little} are theories in which the mass of the
scalar field $\sigma$ (the Higgs) is stabilized against radiative corrections
by making the scalar field a PNGB resulting from a spontaneously
broken (approximate) symmetry ${\cal G}$ at some scale $f$. 
In order to generate a potential for $\sigma$ one needs to explicitly break
the initial global symmetry. The novel feature is that instead of breaking
the initial symmetry with a single coupling, one introduces two couplings
such that each coupling by itself preserves sufficient amount
of symmetry to guarantee the masslessness of the PNGB. Schematically,
to the initial Lagrangian ${\cal L}_0$ one adds two terms 
$\left({\cal L}_1+{\cal L}_2\right)$ with couplings $g_1$ and $g_2$ 
respectively and each term is 
chosen such that by itself it preserves a different subset of global symmetries
under which the $\sigma$ is an exact Nambu-Goldstone boson. The one-loop mass
of $\sigma$ is then necessarily proportional to the product of $g_1
g_2$

\be
m^2\simeq c\,\frac{g_1^2 g_2^2}{16\pi^2}\,f^2.
\ee
where $c$ is a coefficient of order unity 
 which we shall assume is positive.
The Little Higgs potential is not necessarily sinusoidal, but
it is periodic with periodic of order $f$. To obtain order of magnitude
estimates we therefore set $v\sim f$. The estimates will be valid provided
that the curvaton field during inflation is small enough that the 
quadratic approximation \eq{vpot2} is roughly valid.

 Taking for simplicity $g_1\sim g_2\sim g$, the
 decay rate of the curvaton is given by
\be
\Gamma\sim 10^{-1} \frac{m^3}{f^2}\sim 10^{-4}\,g^6\,f.
\ee
The conditions \eqss{bbn}{moverv}{voverplanck} give respectively
\bea
g^6 f & \gsim & 10^{-20}\GeV  \\
g^2 &\lsim&  10^{-1} \\
g &\lsim& 10^2 f/\mpl
\eea
The consistency of the first condition with the third places a lower bound
on the scale of the Little Higgs,
\be
f\gsim 10^{11}\GeV
\,,
\ee
 and the third condition itself requires
that this scale  should be rather high if the coupling $g$ is not 
to be very small. Subject to these conditions, a Little Higgs seems
to be a viable curvaton candidate.

\section{The curvaton as a string axion}

\label{axion}
\noindent
Another interesting possibility is considering the so--called string axion 
fields as potential curvatons. The string axions are the imaginary parts of
the string moduli fields, which correspond to flat directions that are not 
lifted by supergravity corrections. Moduli fields are a necessary ingredient 
of string theories. Their K\"{a}hler potential at tree level is

\begin{equation}
K=-M_P^2\sum_i\ln[(S_i+\bar{S}_i)/M_{\rm P}]\;,
\label{K}
\end{equation}
which is independent from Im$S_i$. Moreover, even though the superpotential 
for the moduli may receive a non-perturbative contribution of the form 
\begin{equation}
W\simeq\sum_j\Lambda_j^3\exp(-\sum_i\beta_{ij}S_i/M_{\rm P})
\label{W}
\end{equation}
the $F$-term scalar potential $V_F$ (which is only a function of $K$ and $W$ 
and their derivatives) may remain flat in the Im$S_i$ directions because,
in general, $W$ is dominated by the term with the largest $\Lambda_j$,
in which case $V_F$ is independent of the phase of $W$. The Im$S_i$ satisfy an 
additional discrete symmetry 

\begin{equation}
{\rm Im} S_i={\rm Im} S_i + 2\pi v\;,
\end{equation}
where, \mbox{$v\lsim M_{\rm P}$}.
The string axions of 4-dimensional supergravity are associated with the 
components of the antisymmetric tensor $B$ of string theory \cite{witten}

\begin{equation}
B=b_{\mu\nu}dx^\mu dx^\nu+b_I\omega^I_{\alpha\beta}dy^{\alpha}dy^{\beta}\;,
\end{equation}
where $\omega^I_{\alpha\beta}$ represents the topology of the compactified 
space corresponding to the extra dimensions~$y^\alpha$. The so-called 
model--independent string axion Im$S$ is related to the four-dimensional 
spacetime components $b_{\mu\nu}$, whereas the model-dependent string axions 
Im$T_i$ are related to $b_I$, which depends on the compactification. 
Their respective real parts are the dilaton \mbox{Re$S=4\pi/g_{\rm GUT}^2$} 
and the so-called geometrical moduli \mbox{Re$T_i$}, associated to the volume 
of the extra dimensions. Thus, string theory provides us with natural 
candidates for the curvaton since the string axions are PNGBs, whose potential 
appears due to SUSY breaking both during and after inflation.

In contrast to the other PNGB cases in the case of the string axion one has
to worry for possible corrections to the potential due to SUSY breaking during
inflation. In view of Eqs.~(\ref{K}) and (\ref{W}) a simple calculation shows 
that the effective mass (due to SUSY breaking) of the canonically normalized 
string axion field is\footnote{This result holds true even when 
$W$ is dominated by more than one terms of the sum in Eq.~(\ref{W}), as is the 
case e.g. of multiple gaugino condensates \cite{decarlos} (racetrack 
scenario \cite{racetrack}).}
\begin{equation}
m_*^2=V_F''(\sigma)\sim\sqrt{V_*}\left(\frac{\Lambda}{M_P}\right)^3\;,
\end{equation}
where \mbox{$\Lambda\equiv$ max$\{\Lambda_j\}$}.
This turns out to be much smaller than $H_*^2$ if \mbox{$V_*^{1/4}\geq M_S$},
where \mbox{$M_S\sim\sqrt{\Lambda^3/M_P}\sim 10^{11}$GeV} is the 
SUSY breaking scale in the vacuum (for gravity mediated SUSY breaking).
Thus, during inflation the string axion is overdamped and remains effectively
frozen. As discussed below, if the string axion is to be a successful curvaton,
it is hard for the inflationary scale to be very low, which results in the 
complete randomization of the field.
Thus, at the end of inflation the expected misalignment of the field is 
maximal, i.e. \mbox{$\sigma_*\sim v$}. After the end of inflation the string 
axion potential is not disturbed by K\"{a}hler corrections, as is the case for 
all PNGBs. It is reasonable to assume, then, that the potential remains 
negligible\footnote{even though it may be possibly modified in early
times by the evolution of Re$T$, while Im$T$ is frozen.}
until much later times when the string axion unfreezes and begins 
its oscillations. At that time the height of the potential is 
given (due to gaugino condensates and also by worldsheet, 
membrane instantons) by the scale of SUSY breaking $M_S$. 

Now, typically, \mbox{$v\sim M_{\rm P}$}, which 
suggests that the mass of the string axion is 
$m\sim m_{3/2}\sim 1$~TeV. One, then, finds 
\mbox{${\cal P}_\zeta^{\frac{1}{2}}\sim 
r{\cal P}_{\delta\sigma/\sigma}^{\frac{1}{2}}\sim rH_*/M_{\rm P}$}, which, 
according to Eq.~(\ref{hoversigma}), gives 
\mbox{$V_*^{1/4}\sim r^{-1/2}10^{16}$GeV}. 
However, this violates the bound of Eq.~(\ref{hbound1}). As a result
the contribution of the inflaton's curvature perturbation is not negligible, 
in contrast to what is usually assumed under the curvaton hypothesis. Thus, we 
see that {\em generically the string axions contribute substantially to the 
curvature perturbation} even under the inflaton hypothesis. This may generate 
important isocurvature perturbations because the inflaton and the string axion 
perturbations are uncorrelated. Also, since \mbox{$v\sim M_P$}, the spectral 
index of the curvaton perturbations is \mbox{$n\approx 1$}. 
Therefore, we expect the spectral index of the overall 
spectrum to be given by the inflaton perturbations on the large (small) scales 
and by the string axion perturbations on small (large) scales if the 
inflaton's perturbation spectrum is red (blue). The switch-over scale is model
dependent and is determined by the relative importance of the contribution to 
the curvature perturbations coming from the inflaton and the string axions 
respectively. Note that, since \mbox{$m\sim 1$ TeV} the string axions violate
the BBN constraint Eq.~(\ref{bbn}). This is the well known moduli problem,
which, however, can be overcome using one of the mechanisms that string 
theory assumes, e.g. a brief period of thermal inflation.

The above scenario is interesting in its own right but it does not benefit
from the liberation effects of the curvaton hypothesis to inflationary model 
building, since the inflaton's curvature perturbations are also important. 
Fortunately, many string theory models offer alternative possibilities. In 
particular, it is possible to have \mbox{$v<M_{\rm P}$} 
\cite{dine,choikim,choi}. Furthermore, strongly 
coupled string theory allows for large values of the dilaton and the geometric
moduli, in which case the scale of $V_F$ is suppressed as
\cite{choi,cck,banks} \mbox{$(vm)^2\sim e^{-2\tau}M_S^4$}, where 
\mbox{$\tau\equiv \pi$ Re$T/M_{\rm P}$} parameterizes the overall modulus $T$. 

Let us investigate the situation by taking\footnote{Note that the effective
mass $m_*$ of the string axion during inflation is also suppressed 
by $e^{-\tau}$.}
\begin{equation}
\sigma_*\sim
v\equiv\alpha M_{\rm P}
\qquad\Rightarrow\qquad m\sim\alpha^{-1}e^{-\tau}m_{3/2}\;,
\label{sax}
\end{equation}
where \mbox{$\alpha\leq 1$} and for gravitational decay of the string axion we 
will use \mbox{$\Gamma\sim m^3/M_{\rm P}^2$}.
The first requirement is that the decay of the string axion occurs before BBN. 
In view of Eq.~(\ref{bbn}), this requirement results in the bound

\begin{equation}
\alpha<
0.1\,e^{-\tau}\;.
\label{bbncons}
\end{equation}

Now, it is easy to see that the COBE requirement 
(c.f. Eq.~(\ref{hoversigma})), in this case, demands
\begin{equation}
V_*^{1/4}\sim\sqrt{\alpha}\,M_{\rm P}{\cal P}_\zeta^{\frac{1}{4}}
r^{-1/2}
\label{V*}
\end{equation}

The dynamics of the string axion depend on whether the field begins to 
oscillate after the onset of radiation domination or not. 
Comparing $H\sub{REH}$ with \mbox{$H_{\rm OSC}\sim m$} we find that 

\begin{equation}
H\sub{REH}\lgeq m\Leftrightarrow
\alpha\lgeq e^{-\tau}\;10^3\gamma^{-1}\;,
\label{begin}
\end{equation}
where we defined 
\begin{equation}
\gamma\equiv \frac{H\sub{REH}}{1\;{\rm GeV}}>10^{-24}
\label{gamma}
\end{equation}
and the lower bound is due to the requirement that the radiation era begins 
before BBN, i.e. \mbox{$H_{\rm BBN}<H\sub{REH}$}. 

>From Eq.~(\ref{begin}) and in view of the constraint (\ref{bbncons}), we see 
that BBN demands that the string axion begins oscillating before reheating if 
\mbox{$\gamma<10^4$}. In the case where the radiation era begins after the 
inflationary reheating we have
\mbox{$H\sub{REH}\sim\Gamma_\Phi$}, where $\Gamma_\Phi$ is the decay rate of 
the inflaton field. Then, the gravitino bound on the reheating temperature 
\mbox{$T_{\rm REH}\sim\sqrt{\Gamma_\Phi M_P}\leq 10^9$GeV} suggests
that \mbox{$\gamma\leq 1$}. However, if the string axion dominates the 
Universe before it decays the gravitino bound may be substantially relaxed by 
the subsequent entropy production \cite{dine} and we may have 
\mbox{$\gamma\gg 1$} without problem. Furthermore, even if the string axion 
decays before it dominates the Universe, one may still have 
\mbox{$\gamma\gg 1$} if extra entropy production occurred during the period 
between the inflationary reheating and the field's decay. This may well be 
possible if, during this period, there was a brief period of thermal inflation 
or the Universe was dominated by another oscillating field, whose curvature 
perturbation (like the inflaton's) is also negligible. Due to the above we 
will treat $\gamma$ as a free parameter. 

\bigskip

\noindent{\bf\boldmath 
1. Onset of curvaton oscillations before radiation domination ($H\sub{REH}<m$)}

\medskip

{\it Case A:} Consider firstly that the curvaton decays before dominating. 
In this case, from Eq.~(\ref{r1}) and using Eq.~(\ref{sax}) we have
\begin{equation}
r\sim\alpha^{7/2}e^{3\tau/2}10^{14}
\gamma^{1/2}\;.
\label{ra}
\end{equation}

Then Eq.~(\ref{V*}) becomes
\begin{equation}
V_*^{1/4}\sim\alpha^{-5/4}e^{-3\tau/4}10^9\gamma^{-1/4}{\rm GeV}\;.
\label{V*1}
\end{equation}
Using the WMAP range \mbox{$10^{-2}\leq r<1$}, Eq.~(\ref{ra}) gives
\begin{equation}
e^{-3\tau/7}10^{-5}\gamma^{-1/7}
\leq\alpha<{\rm min}\left\{e^{-3\tau/7} 10^{-4}\gamma^{-1/7}
,\,e^{-\tau}0.1\right\}\;,
\label{alpha1}
\end{equation}
where we also took Eq.~(\ref{bbncons}) into account. In view of the above, 
Eq.~(\ref{V*1}) becomes
\begin{equation}
\max\left\{e^{-3\tau/14}10^{14}\gamma^{-1/14}, e^{\tau/2}10^{10}
\gamma^{-1/4}\right\}{\rm GeV}< V_*^{1/4}\leq e^{-3\tau/14}10^{15}
\gamma^{-1/14}{\rm GeV}\;,
\label{V*11}
\end{equation}
where the second element in the brackets is due to the BBN constraint.

{\it Case B:}
Now consider that the curvaton dominates before decaying. In this case,
since \mbox{$r\sim 1$}, Eq.~(\ref{V*}) is simply
\begin{equation}
V_*^{1/4}\sim\sqrt{\alpha}\,10^{16}{\rm GeV}\;.
\label{V*2}
\end{equation}
Hence, in view of Eq.~(\ref{bbncons}) and also demanding that
\mbox{$\Gamma<H_{\rm DOM}$} ,
 we find
\begin{equation}
e^{-3\tau/7}10^{-4}\gamma^{-1/7}\leq\alpha<e^{-\tau}0.1\;,
\label{alpha2}
\end{equation}
through which Eq.~(\ref{V*2}) is recast as
\begin{equation}
e^{-3\tau/14}10^{14}\gamma^{-1/14}{\rm GeV}\leq V_*^{1/4}<
e^{-\tau/2}10^{16}{\rm GeV}\;.
\end{equation}

\medskip

\noindent{\bf\boldmath 2. Onset of curvaton oscillations after radiation 
domination ($H\sub{REH}\geq m$)}

\medskip

{\it Case A:} 
Consider that the curvaton decays before dominating. 
In this case Eqs.~(\ref{r1}) and (\ref{sax}) give
\begin{equation}
r\sim\alpha^3e^\tau 10^{15}
\label{rb}\;.
\end{equation}
Then, from Eqs.~(\ref{V*}) and (\ref{rb}), we find:
\begin{equation}
V_*^{1/4}\sim\alpha^{-1}e^{-\tau/2}10^9{\rm GeV}\;.
\label{V*1b}
\end{equation}
Employing again the WMAP range \mbox{$10^{-2}\leq r<1$}, Eq.~(\ref{rb}) gives
\begin{equation}
e^{-\tau/3}10^{-6}\leq\alpha<{\rm min}\left\{e^{-\tau/3} 10^{-5},
\,e^{-\tau}0.1\right\}\;,
\label{alpha1b}
\end{equation}
where we also took Eq.~(\ref{bbncons}) into account. Using the above in 
Eq.~(\ref{V*1b}) we obtain:
\begin{equation}
\max\left\{e^{-\tau/6}10^{14}, e^{\tau/2}10^{10}\right\}{\rm GeV}< 
V_*^{1/4}\leq e^{-\tau/6}10^{15}\;,
\label{V*11b}
\end{equation}
where the second element in the brackets is due to the BBN constraint.

{\it Case B:}
Now consider that the curvaton dominates before decaying. In this case
Eq.~(\ref{V*}) suggests that $V_*^{1/4}$ is again given by Eq.~(\ref{V*2}).
Using the BBN constraint (\ref{bbncons}) and also demanding that
\mbox{$\Gamma<H_{\rm DOM}$},
 we find
\begin{equation}
e^{-\tau/3}10^{-5}\leq\alpha<e^{-\tau}0.1\;,
\label{alpha2b}
\end{equation}
through which Eq.~(\ref{V*2}) is recast as
\begin{equation}
e^{-\tau/6}10^{14}{\rm GeV}\leq V_*^{1/4}<e^{-\tau/2}10^{16}{\rm GeV}\;.
\end{equation}
The above show that as $\alpha$ grows the string axion decays later, 
because \mbox{$\Gamma\propto m^3\propto\alpha^{-3}$}. However, 
Eqs.~(\ref{V*1}), (\ref{V*1b}) and (\ref{V*2}) show that, for growing 
$\alpha$, the energy scale of inflation decreases if the string axion decays 
before domination (Case A) but increases instead if the latter decays after it 
dominates the Universe (Case B). Thus, the minimum possible $V_*$ (for a given 
$\tau$) occurs when the decay of the string axion takes place at just about 
the time when its density comes to dominate the Universe. This value is
\begin{equation}
(V_*^{1/4})_{\rm min}\sim 
\left\{\begin{array}{ll}
e^{-3\tau/14}10^{14}\gamma^{-1/14}\,{\rm GeV} & H_{\rm REH}<m\\
\\
e^{-\tau/6}10^{14}\,{\rm GeV} & H_{\rm REH}\geq m
\end{array}\right.\;.
\end{equation}

In the case when the oscillations begin before [after] the onset of radiation 
domination, from Eq.~(\ref{alpha2}) [Eq.~(\ref{alpha2b})] we see that, 
it is acceptable for the string axion to decay after domination (Case B) only 
if \mbox{$\tau<12+\frac{1}{4}\ln\gamma$} [\mbox{$\tau<14$}].
In this range the BBN constraint is subdominant in the case when the string 
axion decays before domination (Case A). For larger $\tau$ however, it is
the BBN constraint that sets the bounds in Eqs.~(\ref{alpha1}) and 
(\ref{V*11}) [Eqs.~(\ref{alpha1b}) and (\ref{V*11b})]. Still, $\tau$ cannot 
grow much larger. Indeed, the parameter space for subdominant string axion 
decay disappears too when \mbox{$\tau\rightarrow 16+\frac{1}{4}\ln\gamma$}
[\mbox{$\tau\rightarrow 17$}], which is much smaller than the requirements 
of strongly coupled heterotic string theory \cite{choi,cck}: 
\mbox{$\tau\lsim \pi/\alpha_{\rm GUT}\approx 79$} even if $\gamma$ is 
large.\footnote{The maximum value of
$\gamma$ corresponds to prompt inflationary reheating with 
\mbox{$V_*^{1/4}\sim 10^{16}$GeV}, which gives
\mbox{$\gamma_{\rm max}\sim 10^{14}$} and, therefore, 
\mbox{$\tau_{\rm max}\simeq 24$}.}
However, the parameter space can be enlarged toward larger values of $\tau$ 
if the BBN constraint is relaxed, which may be possible if one considers say a 
brief period of thermal inflation subsequent to the string axion decay.

Just to get a feeling 
for our results let us choose \mbox{$\alpha\sim 10^{-2}$} so that 
\mbox{$v\sim 10^{16}$GeV} as suggested also in Refs.~\cite{choikim,choi}. 
Enforcing this into the above one finds that the string axion can indeed act 
as a curvaton but only if we consider weakly coupled string theory, where 
\mbox{$e^\tau\sim 1$} (mainly due to the BBN constraint). Then, for 
\mbox{$\gamma<10^5$}, the string axion begins oscillating before the onset of 
radiation domination. If \mbox{$\gamma<10^{-14}$} the field decays before it 
dominates the Universe (Case~A), in which case the COBE requirement demands 
\mbox{$V_*^{1/4}\sim 10^{11}\gamma^{-1/4}$GeV}. If, however, 
\mbox{$10^{-14}\leq\gamma<10^5$} the field manages to dominate the Universe
before decaying (Case~B) and \mbox{$V_*^{1/4}\sim 10^{15}$GeV}. For 
\mbox{$\gamma\geq 10^5$} the oscillations of the string axion begin after the
onset of radiation domination. In this case the curvaton requirements are
impossible to satisfy if the string axion decays before domination (Case~A)
and only the case when the string axion dominates the Universe (Case~B) 
is allowed, which, again, demands \mbox{$V_*^{1/4}\sim 10^{15}$GeV}.
Thus, we see that with \mbox{$\alpha\sim 10^{-2}$} it is marginally possible 
to liberate inflation from the COBE constraint (c.f. Eq.~(\ref{hbound1})). 
Better results are obtained for smaller values of $\alpha$, which also ensure 
the protection of the flatness of the potential against quantum gravity 
effects since \mbox{$v\ll M_{\rm P}$}. For example, for 
\mbox{$\alpha\sim 10^{-4}$} one finds \mbox{$V_*^{1/4}\sim 10^{14}$GeV} with 
\mbox{$\tau\leq 9$}.

In summary we have shown that a string axion can be a successful curvaton
but one needs to reduce the inflationary scale enough so that the contribution 
of the inflaton to the overall curvature perturbation is negligible. This is 
possible only if $v$ is substantially smaller than $M_P$, which also increases 
the axion mass over 1~TeV and solves the moduli problem \cite{dine}. A value 
\mbox{$v\sim 10^{16}$GeV} or smaller is not unreasonable in some string theory 
models \cite{choikim,choi}. The moduli problem reappears if one considers 
strongly coupled string theory (M-theory), where BBN is challenged again for
large values of Re$T$.

\section{Conclusion}

\label{conclusion}
\noindent

Mainly motivated by the fact that  supersymmetry is badly broken 
during inflation 
and therefore of limited  use in  keeping  scalar fields light, in this paper 
we have analyzed the possibility that cosmological perturbations
are generated by a curvaton field and that the latter is 
a PNGB. In such  a case, the mass of the curvaton field
during inflation is protected by some symmetry and -- in contrast to 
inflationary models where the inflaton field is a PNGB --
the vanishing of the curvaton potential in the limit
of exact symmetry does not pose any problem.
We have given a general framework for discussing PNGB curvaton candidates,
and explored different options. 

We have
shown that the curvaton may be identified with the
fifth component of  a gauge field living in five dimensions. The
finiteness of the potential is provided by gauge invariance in five
dimensions and no supersymmetry is required. It turns out that
the size of the extra-dimension has to be smaller than about $10^8$
of the planckian length scale for the scenario to be viable.

Alternatively, the curvaton mass may be kept light within
Little Higgs theories in which the mass of the
scalar field is stabilized against radiative corrections
by making the scalar field a PNGB resulting from a spontaneously
broken (approximate) symmetry ${\cal G}$ at some scale $f$.
The  novel feature is that instead of breaking
the initial symmetry with a single coupling, one introduces two couplings
such that each coupling by itself preserves sufficient amount
of symmetry to guarantee the masslessness of the PNGB. 
We have found that the scale $f$ must exceed $10^{11}\GeV$ for this 
to work, and that to avoid a very small coupling the scale should be
considerably higher.


Finally, we have thoroughly
studied the case in which the curvaton is the string axion, the imaginary part 
of the string moduli fields corresponding to flat directions that are not 
lifted by supergravity corrections. In this case we have shown that 
supersymmetry breaking during inflation does not lift the flatness of the
potential if the inflationary energy scale is larger than the scale of
supersymmetry breaking in the vacuum. Then the string axion may be a successful
curvaton if the order parameter $v$ is smaller than $M_P$ so that the 
inflaton's contribution to the curvature perturbations is negligible.
Overall, our findings indicate that
requiring  that the effective curvaton  mass be much less than the Hubble
parameter during inflation is a severe constraint, but can be 
successfully satisfied if the curvaton is a PNGB. 

 \subsection*{Acknowledgments}
This work was supported by PPARC and by the EU Fifth Framework network 
"Supersymmetry and the Early Universe" HPRN-CT-2000-00152.

 \newcommand\pl[3]{Phys.\ Lett.\ {\bf #1}  (#3) #2}
 \newcommand\np[3]{Nucl.\ Phys.\ {\bf #1}  (#3) #2}
 \newcommand\pr[3]{Phys.\ Rep.\ {\bf #1}  (#3) #2}
 \newcommand\prl[3]{Phys.\ Rev.\ Lett.\ {\bf #1}  (#3)  #2}
 \newcommand\prd[3]{Phys.\ Rev.\ D{\bf #1}  (#3) #2}
 \newcommand\ptp[3]{Prog.\ Theor.\ Phys.\ {\bf #1}  (#3)  #2 }
 \newcommand\rpp[3]{Rep.\ on Prog.\ in Phys.\ {\bf #1} (#3) #2}
 \newcommand\jhep[2]{JHEP #1 (#2)}
 \newcommand\grg[3]{Gen.\ Rel.\ Grav.\ {\bf #1}  (#3) #2}
 \newcommand\mnras[3]{MNRAS {\bf #1}   (#3) #2}
 \newcommand\apjl[3]{Astrophys.\ J.\ Lett.\ {\bf #1}  (#3) #2}


\begin{thebibliography}{99}

\bibitem{treview} For a review, see  D.~H.~Lyth and A.~Riotto,
Phys.\ Rept.\  {\bf 314}, 1 (1999); 
A.~Riotto,
arXiv:hep-ph/0210162.

\bibitem{sylvia} 
S.~Mollerach,
Phys.\ Rev.\ D {\bf 42}, 313 (1990).

\bibitem{lm}
A.~D.~Linde and V.~Mukhanov,
Phys.\ Rev.\ D {\bf 56}, 535 (1997).

\bibitem{lw}
D.~H.~Lyth and D.~Wands,
Phys.\ Lett.\ B {\bf 524}, 5 (2002).


\bibitem{mt1} 
T.~Moroi and T.~Takahashi,
Phys.\ Lett.\ B {\bf 522}, 215 (2001).

\bibitem{andrew} 
N.~Bartolo and A.~R.~Liddle,
Phys.\ Rev.\ D {\bf 65}, 121301 (2002).

\bibitem{mt2} 
T.~Moroi and T.~Takahashi,
arXiv:hep-ph/0206026.

\bibitem{fy}
M.~Fujii and T.~Yanagida,
Phys.\ Rev.\ D {\bf 66}, 123515 (2002)
[arXiv:hep-ph/0207339].

\bibitem{luw}
D.~H.~Lyth, C.~Ungarelli and D.~Wands,
arXiv:astro-ph/0208055.

\bibitem{hmy}
A.~Hebecker, J.~March-Russell and T.~Yanagida,
arXiv:hep-ph/0208249.

\bibitem{hofmann} 
R.~Hofmann,
arXiv:hep-ph/0208267.

\bibitem{dl}
K.~Dimopoulos and D.~H.~Lyth,
arXiv:hep-ph/0209180.

\bibitem{bck1} 
M.~Bastero-Gil, V.~Di Clemente and S.~F.~King,
arXiv:hep-ph/0211011.

\bibitem{bck2} 
M.~Bastero-Gil, V.~Di Clemente and S.~F.~King,
arXiv:hep-ph/0211012.

\bibitem{mormur}
T.~Moroi and H.~Murayama,
arXiv:hep-ph/0211019.

\bibitem{ekm} 
K.~Enqvist, S.~Kasuya and A.~Mazumdar,
arXiv:hep-ph/0211147.

\bibitem{mwu} 
K.~A.~Malik, D.~Wands and C.~Ungarelli,
arXiv:astro-ph/0211602.

\bibitem{postma} 
M.~Postma,
arXiv:hep-ph/0212005.

\bibitem{fl}
B.~Feng and M.~Li, arXiv:hep-ph/0212213.

\bibitem{gl}
C.~Gordon and A.~Lewis,
arXiv:astro-ph/0212248.

\bibitem{kostasquint}
K.~Dimopoulos,
arXiv:astro-ph/0212264.

\bibitem{giov}
M.~Giovannini,
arXiv:hep-ph/0301264.

\bibitem{lu}
A.~R.~Liddle and L.~A.~Urena-Lopez,
arXiv:astro-ph/0302054.

\bibitem{omit}
J.~McDonald,
arXiv:hep-ph/0302222.

\bibitem{dllr1}
K.~Dimopoulos, G.~Lazarides, D.~Lyth and R.~Ruiz de Austri,
arXiv:hep-ph/0303154.

\bibitem{ejkm}
K.~Enqvist, A.~Jokinen, S.~Kasuya and A.~Mazumdar,
arXiv:hep-ph/0303165.

\bibitem{03dl}
D.\ H.\ Lyth \& D.\ Wands, in preparation.


\bibitem{dllr2}
K.~Dimopoulos, G.~Lazarides, D.~Lyth and R.~R.~de Austri,
arXiv:hep-ph/0308015.

\bibitem{dgz}
G.~Dvali, A.~Gruzinov and M.~Zaldarriaga,
arXiv:astro-ph/0303591.

\bibitem{kofman}
L.~Kofman,
arXiv:astro-ph/0303614.

\bibitem{dfn}
M.~Dine, W.~Fischler and D.~Nemeschansky,
Phys.\ Lett.\ B {\bf 136}, 169 (1984).

\bibitem{chrr}
G.~D.~Coughlan, R.~Holman, P.~Ramond and G.~G.~Ross,
Phys.\ Lett.\ B {\bf 140}, 44 (1984).

\bibitem{cllsw}
E.~J.~Copeland, A.~R.~Liddle, D.~H.~Lyth, E.~D.~Stewart and D.~Wands,
Phys.\ Rev.\ D {\bf 49}, 6410 (1994)
[arXiv:astro-ph/9401011].

\bibitem{drt}
M.~Dine, L.~Randall and S.~Thomas,
Nucl.\ Phys.\ B {\bf 458}, 291 (1996)
[arXiv:hep-ph/9507453].

\bibitem{pre} G.~W.~Anderson, A.~D.~Linde and A.~Riotto,
Phys.\ Rev.\ Lett.\  {\bf 77}, 3716 (1996)
[arXiv:hep-ph/9606416]; G.~R.~Dvali and A.~Riotto,
Phys.\ Lett.\ B {\bf 388}, 247 (1996)
[arXiv:hep-ph/9606431]; A.~Riotto, E.~Roulet and I.~Vilja,
Phys.\ Lett.\ B {\bf 390}, 73 (1997)
[arXiv:hep-ph/9607403].

\bibitem{ewan}
E.~D.~Stewart,
Phys.\ Rev.\ D {\bf 51}, 6847 (1995)
[arXiv:hep-ph/9405389].

\bibitem{lmg}
M.~K.~Gaillard, D.~H.~Lyth and H.~Murayama,
Phys.\ Rev.\ D {\bf 58}, 123505 (1998)
[arXiv:hep-th/9806157].

\bibitem{cr}
J.~A.~Casas, G.~B.~Gelmini and A.~Riotto,
Phys.\ Lett.\ B {\bf 459}, 91 (1999)
[arXiv:hep-ph/9903492].

\bibitem{subir}
J.~A.~Adams, G.~G.~Ross and S.~Sarkar,
Phys.\ Lett.\ B {\bf 391} (1997) 271
[arXiv:hep-ph/9608336].

\bibitem{cs}
J.~D.~Cohn and E.~D.~Stewart,
Phys.\ Lett.\ B {\bf 475}, 231 (2000)
[arXiv:hep-ph/0001333];
E.~D.~Stewart and J.~D.~Cohn,
Phys.\ Rev.\ D {\bf 63}, 083519 (2001)
[arXiv:hep-ph/0002214].

\bibitem{newwilson1} N.~Arkani-Hamed, 
H.~C.~Cheng, P.~Creminelli and L.~Randall,
hep-th/0301218.

\bibitem{newwilson2} D.~E.~Kaplan and N.~Weiner, hep-ph/0302014.

\bibitem{newwilson3}N.~Arkani-Hamed, 
H.~C.~Cheng, P.~Creminelli and L.~Randall,   hep-th/0302034.

\bibitem{myaxion}
D.~H.~Lyth,
Phys.\ Rev.\ D {\bf 45}, 3394 (1992).

\bibitem{LLbook}
A.~R.~Liddle and D.~H.~Lyth,
``Cosmological inflation and large-scale structure,''
{\it  Cambridge, UK: Univ. Pr. (2000) 400 p}.

\bibitem{wmapparam}
D.~N.~Spergel {\it et al.}, astro-ph/0302209.

\bibitem{wmapnong} 
E.~Komatsu {\it et al.}, astro-ph/0302223.

\bibitem{LL1}
A.~R.~Liddle and D.~H.~Lyth,
Phys.\ Lett.\ B {\bf 291}, 391 (1992)
[arXiv:astro-ph/9208007].

\bibitem{nongauss} V.~Acquaviva, N.~Bartolo, S.~Matarrese and A.~Riotto,
arXiv:astro-ph/0209156; J.~Maldacena,
arXiv:astro-ph/0210603.

\bibitem{martin}
M.~Tegmark and M.~J.~Rees,
Astrophys.\ J.\  {\bf 499}, 526 (1998)
[arXiv:astro-ph/9709058].

\bibitem{prr} L.~Pilo, D.~A.~Rayner and A.~Riotto,
hep-ph/0302087.

\bibitem{hosotani}
Y.~Hosotani,
Phys.\ Lett.\ B {\bf 126}, 309 (1983); {\it ibidem} 
Phys.\ Lett.\ B {\bf 129}, 193 (1983); {\it ibidem}
Annals Phys.\  {\bf 190}, 233 (1989).

\bibitem{pq} A.~Delgado, A.~Pomarol and M.~Quiros,
Phys.\ Rev.\ D {\bf 60}, 095008 (1999).

\bibitem{little} N.~Arkani-Hamed, A.~G.~Cohen, E.~Katz and A.~E.~Nelson,
JHEP {\bf 0207}, 034 (2002); 
N.~Arkani-Hamed, A.~G.~Cohen, 
E.~Katz, A.~E.~Nelson, T.~Gregoire and J.~G.~Wacker,
JHEP {\bf 0208}, 021 (2002); 
T.~Gregoire and J.~G.~Wacker,
JHEP {\bf 0208}, 019 (2002); T.~Gregoire and J.~G.~Wacker, hep-ph/0207164; 
I.~Low, W.~Skiba and D.~Smith,
Phys.\ Rev.\ D {\bf 66}, 072001 (2002);
D.~E.~Kaplan and M.~Schmaltz,
hep-ph/0302049.

\bibitem{witten}
E.~Witten,
Phys.\ Lett.\ B {\bf 149}, 351 (1984).

\bibitem{decarlos}
T.~Barreiro and B.~de Carlos,
JHEP {\bf 0003}, 020 (2000)
[arXiv:hep-ph/9912387].

\bibitem{racetrack}
B.~de Carlos, J.~A.~Casas and C.~Munoz,
Nucl.\ Phys.\ B {\bf 399}, 623 (1993)
[arXiv:hep-th/9204012].

\bibitem{dine}
T.~Banks and M.~Dine,
Nucl.\ Phys.\ B {\bf 505}, 445 (1997)
[arXiv:hep-th/9608197].

\bibitem{choikim}
K.~Choi and J.~E.~Kim,
Phys.\ Lett.\ B {\bf 165}, 71 (1985);
K.~Choi and J.~E.~Kim,
Phys.\ Lett.\ B {\bf 154}, 393 (1985)
[Erratum-ibid.\  {\bf 156B}, 452 (1985)].

\bibitem{choi}
K.~Choi,
Phys.\ Rev.\ D {\bf 56}, 6588 (1997)
[arXiv:hep-th/9706171].

\bibitem{cck}
K.~Choi, E.~J.~Chun and H.~B.~Kim,
Phys.\ Rev.\ D {\bf 58}, 046003 (1998)
[arXiv:hep-ph/9801280].

\bibitem{banks}
T.~Banks and M.~Dine,
arXiv:hep-th/9609046.
































 \end{thebibliography}
 \end{document}